\documentclass[a4paper,superscriptaddress,twocolumn,amsmath,nofootinbib]{revtex4}
\usepackage{amsfonts}
\usepackage{graphicx}
\usepackage[latin1]{inputenc}
\usepackage{amsmath}
\usepackage{amssymb}

\begin{document}


\title{Frame-dragging effect in the field of non rotating body due to unit gravimagnetic moment}

\author{Alexei A. Deriglazov }
\email{alexei.deriglazov@ufjf.edu.br} \affiliation{Depto. de Matem\'atica, ICE, Universidade Federal de Juiz de Fora,
MG, Brazil}


\author{Walberto Guzm\'an Ram\'irez }
\email{wguzman@cbpf.br} \affiliation{Depto. de Matem\'atica, ICE, Universidade Federal de Juiz de Fora, MG, Brazil}

\date{\today}

\begin{abstract}
Nonminimal spin-gravity interaction through unit gravimagnetic moment leads to modified
Mathisson-Papapetrou-Tulczyjew-Dixon equations with improved behavior in the ultrarelativistic limit. We present exact
Hamiltonian of the resulting theory and compute an effective $\frac{1}{c^2}$\,-Hamiltonian and leading post-Newtonian
corrections to the trajectory and spin. Gravimagnetic moment causes the same precession of spin ${\bf S}$ as a fictitious rotation
of the central body with angular momentum ${\bf J}=\frac{M}{m}{\bf S}$. So the modified equations imply a number of
qualitatively new effects, that could be used to test experimentally, whether a rotating body in general relativity has
null or unit gravimagnetic moment.
\end{abstract}

\maketitle 


The manifestly generally covariant Mathisson-Papapetrou-Tulczyjew-Dixon (MPTD) equations \cite{Mathisson:1937zz,
Fock1939, Papapetrou:1951pa, Tulc, Dixon1964, pirani:1956} are widely used in general
relativity to describe a rotating test body in pole-dipole approximation. In the current literature (see \cite{Pomeransky1998, Will2014, Gal2013,
Huil2017, Sudal2017, Haroon2017} and references therein), they usually appear in the form given by Dixon
\begin{eqnarray}\label{sh1}
\nabla P_\mu =-\frac{1}{4}\theta_{\mu\nu}\dot
x^\nu \, ,  \qquad
\nabla S^{\mu\nu} = 0 \,,
\end{eqnarray}
where $\theta_{\mu\nu}=R_{\mu\nu\alpha\beta}S^{\alpha\beta}$ is the gravitational analogy of the electromagnetic field
strength $F_{\mu\nu}$ \cite{Pomeransky1998, DWGR2016}. (Our spin-tensor $S^{\mu\nu}$ is twice of that of Dixon.
Besides, in the last equation we omitted the term $2P^{[\mu}\dot x^{\nu]}$, which does not contribute
in $\frac{1}{c^2}$\,-approximation we are interested in the present work. Concerning other notation, see the 
footnote\footnote{Our variables are taken in arbitrary parametrization $\tau$, then $\dot x^\mu=\frac{dx^\mu}{d\tau}$ and the
covariant derivative is $\nabla\omega^\mu=\frac{d\omega^\mu}{d\tau}+\Gamma^\mu_{\alpha\beta}\dot x^\alpha
\omega^\beta$. The square brackets mean antisymmetrization,
$\omega^{[\mu}\pi^{\nu]}=\omega^\mu\pi^\nu-\omega^\nu\pi^\mu$. We often miss the four-dimensional indexes and use the
notation  $\dot x^\mu N_{\mu\nu}\dot x^\nu=\dot xN\dot x$,  $N^\mu{}_\nu\dot x^\nu=(N\dot x)^\mu$,
$\omega^2=g_{\mu\nu}\omega^\mu\omega^\nu$, $\mu, \nu=0, 1, 2, 3$, $\mbox{sign} \, g_{\mu\nu}=(-, +, +, +)$. Suppressing
the indexes of three-dimensional quantities, we use bold letters. The tensor of Riemann curvature is
$R^\sigma{}_{\lambda\mu\nu}=\partial_\mu\Gamma^\sigma{}_{\lambda\nu} -\partial_\nu
\Gamma^\sigma{}_{\lambda\mu}+\Gamma^\sigma{}_{\beta\mu}\Gamma^{\beta}{}_{\lambda\nu}-
\Gamma^\sigma{}_{\beta\nu}\Gamma^{\beta}{}_{\lambda\mu}$.}.) Together with the spin supplementary 
condition\footnote{While the Lagrangian and Hamiltonian formalisms dictate
\cite{hanson1974} the condition (\ref{sh1.2}), in the multipole approach there is a freedom in the choice of a spin
supplementary condition, related with the freedom to choose a representative point of the body \cite{Papapetrou:1951pa,
Tulc, pirani:1956}. Different conditions lead to the same results for observables in $\frac{1}{c^2}$\,-approximation,
see \cite{Schiff1960.1, Dixon1964, Connell1975}. }
\begin{eqnarray}\label{sh1.2}
S^{\mu\nu}P_\nu=0,
\end{eqnarray}
MPTD equations prescribe the evolution of both trajectory and spin of the body in $1/c^2$\,-approximation.

Starting from the pioneer works, MPTD equations were considered as a Hamiltonian-type system. Following this spirit in
the recent work \cite{DW2015.1}, we explicitly realized this idea by constructing the minimal interaction os spin with gravity
in the Lagrangian of vector model of spinning particle, and showed that this indeed leads to MPTD equations in the
Hamiltonian formalism. This allowed us to study ultrarelativistic limit in exact equations for the trajectory of MPTD
particle in the laboratory time.  Using the Landau-Lifshitz $(1+3)$\,-decomposition \cite{bib16} we  observed that,
unlike a geodesic equation, the MPTD equations lead to the expression for three-acceleration which contains divergent
terms as $v\rightarrow c$ \cite{DWGR2016}.  Therefore it seems interesting to find a generalization of MPTD equations
with improved behavior in the ultrarelativistic regime. This can be achieved, if we add a nonminimal spin-gravity
interaction through gravimagnetic moment $\kappa$ \cite{DWGR2016}. $\kappa=0$ corresponds to the MPTD equations. The
most interesting case turns out to be $\kappa=1$ (gravimagnetic body). Keeping only the terms, which may contribute in
the leading post-Newtonian approximation, this gives the modified equations (among other equations, see below)
\begin{eqnarray}\label{sh2}
\nabla P_\mu = -\frac 14 \theta_{\mu\nu}\dot x^\nu -\frac{\sqrt{-\dot x^2}}{32mc} (\nabla_\mu  \theta_{\sigma\lambda})
S^{\sigma\lambda}  \, , \cr \nabla S^{\mu\nu} =\frac{\sqrt{-\dot x^2}}{4mc}\theta^{[\mu}{}_{\alpha} S^{\nu]\alpha}  \,.
\qquad \qquad
\end{eqnarray}
Comparing (\ref{sh2}) with (\ref{sh1}), we see that unit gravimagnetic moment yields quadratic in spin corrections to MPTD equations
in $\frac{1}{c^2}$\,-approximation.

Both acceleration and spin torque of gravimagnetic body have reasonable behavior in ultrarelativistic limit
\cite{DWGR2016}. In the present work we study the modified equations and the corresponding effective Hamiltonian in the
regime of small velocities, and compute $\frac{1}{c^2}$\,-corrections due to the extra-terms appeared in (\ref{sh2}).
In Schwarzschild and Kerr space-times, the modified equations predict a number of qualitatively new effects, that could
be used to test experimentally, whether a rotating body in general relativity has null or unit gravimagnetic moment.

Let us briefly describe the variational problem which implies the modified equations (\ref{sh2}). In the vector model
of spin presented in \cite{deriglazov2014Monster}, the configuration space consist of the position of the particle
$x^\mu(\tau)$, and the vector $\omega^\mu(\tau)$ attached to the point $x^\mu(\tau)$. Minimal interaction with gravity
is achieved by direct covariantization of the free action, initially formulated in Minkowski space. That is we replace
$\eta_{\mu\nu}\rightarrow g_{\mu\nu}$, and usual derivative of the vector $\omega^\mu$ by the covariant derivative:
$\dot\omega^\mu\rightarrow\nabla\omega^\mu$. The nonminimal spin-gravity interaction through the gravimagnetic moment
$\kappa$ can be thought as a deformation of original metric: $g^{\mu\nu}\rightarrow\sigma^{\mu\nu}=g^{\mu\nu}+\kappa
R_\alpha{}^\mu{}_\beta{}^\nu\omega^\alpha\omega^\beta$, with the  resulting Lagrangian action  \cite{DWGR2016}
\begin{eqnarray}\label{sh3}
S=-\int d\tau \sqrt{(mc)^2 -\frac{\alpha}{\omega^2}} ~  \times  \qquad \qquad \cr \sqrt{-\dot xNK\sigma N\dot
x-\nabla\omega NKN\nabla\omega+2\lambda\dot xNKN\nabla\omega}.
\end{eqnarray}
We have denoted $K=(\sigma-\lambda^2g)^{-1}$, where $\lambda$ is the only Lagrangian multiplier in the theory.  The
matrix $N_{\mu\nu}\equiv  g_{\mu\nu}-\frac{\omega_\mu \omega_\nu}{\omega^2}$ is a projector on the plane orthogonal to
$\omega$: $N_{\mu\nu}\omega^\nu=0$. The parameter $\alpha$ determines the value of spin, in particular,
$\alpha=\frac{3\hbar^2}{4}$ corresponds to the spin one-half particle. In the spinless limit, $\omega^\mu=0$ and
$\alpha=0$, Eq. (\ref{sh3}) reduces to the standard Lagrangian of a point particle, $-mc\sqrt{-g_{\mu\nu}\dot x^\mu\dot
x^\nu}$.

The action (\ref{sh3}) is manifestly invariant under general-coordinate transformations as well as under
reparametrizations of the evolution parameter $\tau$. Besides, there is one more local symmetry, which acts in
 the spin-sector and called the spin-plane symmetry: the action remains invariant under rotations of the vectors
$\omega^\mu$ and $\pi_\mu=\frac{\partial L}{\partial\dot\omega^\mu}$ in their own plane \cite{DPM1}. Being affected by
the local transformation, these vectors do not represent observable quantities. But their combination
\begin{eqnarray}\label{sh4}
S^{\mu\nu}=2(\omega^\mu\pi^\nu-\omega^\nu\pi^\mu)=(S^{i0}=D^i, ~ S_{ij}=2\epsilon_{ijk}S_k), ~
\end{eqnarray}
is an invariant quantity, which represents the spin-tensor of the particle. In Eq. (\ref{sh4}), we decomposed the
spin-tensor  into three-dimensional spin-vector ${\bf S}=\frac{1}{2}(S^{23}, S^{31}, S^{12})$, and dipole electric
moment \cite{ba1} $D^i$.

For the general-covariant and spin-plane invariant variables $x^\mu$,
$P_\mu=p_\mu-\Gamma^\beta_{\alpha\mu}\omega^\alpha\pi_\beta$ and $S^{\mu\nu}$  (here $p_\mu=\frac{\delta S}{\delta\dot
x^\mu}$), the Hamiltonian equations of motion of the theory (\ref{sh3}) acquire especially simple form when $\kappa=1$.
In $\frac{1}{c^2}$\,-approximation, we obtained the equations (\ref{sh2}), accompanied by the Hamiltonian equation for
$x^\mu$, $\dot x^\mu=\frac{\sqrt{-\dot x^2}}{mc}P_\mu$, the latter can be identified with velocity-momentum relation
implied by MPTD equations \cite{DWGR2016}.  Besides the dynamical equations, the variational problem (\ref{sh3})
implies the mass-shell constraint
\begin{eqnarray}\label{sh5}
T\equiv P^2+\frac{\kappa}{16}\theta_{\mu\nu}S^{\mu\nu}+(mc)^2=0,
\end{eqnarray}
and the spin-sector constraints  $P\omega =0$ , $P\pi =0$, $\omega\pi=0$ and $\pi^2 - \frac{\alpha}{\omega^2}= 0$.
Their meaning becomes clear if we consider their effect over the spin-tensor. The second-class constraints $P\omega =0$
and  $P\pi =0$ imply the spin supplementary condition (\ref{sh1.2}),  while the remaining first-class constraints fix
the value of square of the spin-tensor, $S^{\mu\nu} S_{\mu\nu} = 8\alpha$. The equations imply that only two components
of spin-tensor are independent, as it should be for an elementary spin one-half particle. The mass-shell constraint
(\ref{sh5}) look like that of a spinning particle with gyromagnetic ratio $g$,
$P^2-\frac{eg}{c}F_{\mu\nu}S^{\mu\nu}+(mc)^2=0$. In view of this similarity, the interaction constant $\kappa$ has been
called gravimagnetic moment \cite{Khriplovich1989, Pomeransky1998}.

Although the vector model of spin has been initially developed to describe an elementary particle of spin one-half, it
can be adopted to study a rotating body in general relativity. The action (\ref{sh3}) with $\kappa=0$ implies MPTD
equations, and the only difference among two formalisms is that values of momentum and spin are conserved quantities of
MPTD equations, while in the vector model they are fixed by constraints. In summary \cite{DWGR2016}, to study the class
of trajectories of a body with $\sqrt{-P^2}=k$ and $S^2=\beta$, we can use our spinning particle with $m=\frac{ k}{c}$
and $\alpha=\frac{\beta}{8}$.

Although the post-Newtonian approximation can be obtained by direct computations on the base of equations of motion, we
prefer to work with an approximate Hamiltonian. This gives  a more transparent picture of nonminimal interaction, in
particular, display strong analogy with a spinning particle with magnetic moment in electromagnetic background. We could
consider a Hamiltonian corresponding to either Poisson or Dirac brackets. We work with Dirac bracket\footnote{The Dirac bracket turns the spinning particle into intrinsically noncommutative theory. This could manifest itself in various 
applications \cite{KaiMa1, KaiMa2, KaiMa3}. In particular, our Hamiltonian differs from those suggested by other groups, for instance \cite{Barausse2009}. They have been compared in \cite{DW2015.1}.} for the
second-class constraints $P \omega =0$ and $P\pi =0$, since in this case the relativistic Hamiltonian acquires the
conventional form $H_{rel}=\frac{\lambda}{2}T$. According to the procedure described in \cite{DPM2016}, exact
Hamiltonian for dynamical variables ${\bf x}(t)$, ${\bf p}(t)$ and ${\bf S}(t)$ as functions of the coordinate time
$t=\frac{x^0}{c}$ is $H=-cp_0$, where the conjugated momentum $p_0$ is a  solution to  the mass-shell constraint
(\ref{sh5}). Solving the constraint, we obtain
\begin{eqnarray}\label{sh7}
H=\frac{c}{\sqrt{-g^{00}}}\sqrt{(mc)^2+\gamma^{ij}P_i P_j+\frac{1}{16}(\theta S)}- \cr
c\pi_\mu\Gamma^\mu{}_{0\nu}\omega^\nu+\frac{cg^{0i}}{g^{00}}P_i,  \qquad \qquad
\end{eqnarray}
where $\gamma^{ij}=g^{ij}-\frac{g^{0i}g^{0j}}{g^{00}}$. Let us consider a stationary, asymptotically flat metric of a
spherical body with mass $M$ and angular momentum ${\bf J}$.  In the post-Newtonian approximation\footnote{We omitted $\frac{1}{c^4}$\,-term in $g_{ij}$ as it does not contribute into the quantities under interest in $\frac{1}{c^2}$\,-approximation.} 
$\frac{1}{c^4}$, this reads \cite{Weinberg, Wald1972}
\begin{eqnarray}\label{sh8}
ds^2 = \left(- 1 +\frac{2GM}{c^2r}-\frac{2G^2M^2}{c^4r^2} \right)(dx^0)^2 - \qquad  \cr 4G\frac{\epsilon_{ijk}J^j x^k}{c^3r^3} dx^0
dx^i +\left( 1+ \frac{2GM}{c^2r}\right)dx^idx^i .
\end{eqnarray}
To obtain the effective Hamiltonian, we expand all quantities in (\ref{sh7}) in series up to $\frac{1}{c^2}$\,-order.
To write the result in a compact form, we introduce the vector potential ${\bf A}_J=\frac{2G}{c}[{\bf
J}\times\frac{{\bf r}}{r^3}]$ for the gravitomagnetic field ${\bf B}_J=[{\boldsymbol{\nabla}}\times {\bf A}_J]=\frac
{2G}{c} \frac{3({\bf J}\cdot \hat{\bf r}) \hat{\bf r}-{\bf J}}{r^3}$, produced by rotation of central body  (we use the
conventional  factor $\frac{2G}{c}$, different from that of Wald \cite{Wald1972}. In the result, our ${\bf B}_J=4{\bf
B}_{Wald}$). Besides we define the vector potential ${\bf A}_S=\frac{M}{m}\frac{G}{c}[{\bf S}\times\frac{{\bf
r}}{r^3}]$ of fictitious gravitomagnetic field ${\bf B}_S=[{\boldsymbol{\nabla}}\times {\bf A}_S]=\frac{M}{m}\frac
{G}{c} \frac{3({\bf S}\cdot \hat{\bf r}) \hat{\bf r}-{\bf S}}{r^3}$ due to rotation of a gyroscope,  as well as the
extended momentum
${\boldsymbol{\Pi}}\equiv{\bf p}+\frac{m}{c}({\bf A}_J+2{\bf A}_S)$.
With these notation, $\frac{1}{c^2}$\,-Hamiltonian becomes similar to that of spinning particle in a magnetic field
\footnote{Here the square root should be expanded up to $\frac{1}{c^2}$\,-order.}
\begin{eqnarray}\label{sh9}
H= \frac{c}{\sqrt{-g^{00}}}\sqrt{(mc)^2+g^{ij}\Pi_i \Pi_j}+\frac{1}{2c}({\bf B}_J+{\bf B}_S)\cdot {\bf S}. ~
\end{eqnarray}
Note that the Hamiltonian $\frac{c}{\sqrt{-g^{00}}}\sqrt{(mc)^2+g^{ij}p_ip_j}$ corresponds to the usual Lagrangian
$L=-mc\sqrt{-g_{\mu\nu}\dot x^\mu\dot x^\nu}$ describing a spinless particle propagating in the Schwarzschild metric
$g_{\mu\nu}$. So, the approximate Hamiltonian (\ref{sh9}) can be thought as  describing a gyroscope orbiting in the
field of Schwarzschild space-time and interacting with the gravitomagnetic field ${\bf B}_J+{\bf B}_S$.

Effective Hamiltonian for MPTD equations turns out to be less symmetric: it is obtained from (\ref{sh9}) excluding the
term $\frac{1}{2c}({\bf B}_S\cdot {\bf S})$, while keeping the potential ${\bf A}_S$ in ${\boldsymbol{\Pi}}$. Hence the
only effect of nonminimal interaction is the deformation of gravitomagnetic field of central body according to the
rule
\begin{eqnarray}\label{sh10}
{\bf B}_J\rightarrow{\bf B}_J+{\bf B}_S.
\end{eqnarray}
Eq. (\ref{sh9}) together with the Dirac brackets, also taken in $\frac{1}{c^2}$\,-approximation, gives us Hamiltonian
equations of motion for ${\bf x}(t)$, ${\bf p}(t)$ and ${\bf S}(t)$.  Excluding from them the momentum ${\bf p}$, we
obtain acceleration and spin precession of gravimagnetic particle in $\frac{1}{c^2}$\,-approximation.

Total acceleration of gravimagnetic particle in $\frac{1}{c^2}$\,-approximation reads (here $\hat{\bf r}={\bf r}/|{\bf r}|$)
\begin{eqnarray}\label{sh11}
\mathbf{a}=  -\frac{MG}{r^2} \mathbf{\hat r} + \frac{4GM}{c^2r^2}(  \hat{\mathbf{r}}\cdot\mathbf{v} )
\mathbf{v} -\frac{GM}{c^2r^2} v^2\mathbf{\hat r} + \frac{4G^2M^2}{c^2r^3}\mathbf{\hat r} +  \cr
\frac{1}{c}\left({\bf B}_J+{\bf B}_S\right)\times{\bf v}+
\frac{GM}{mc^2r^3} \left[ {\bf S}\times{\bf v}+3({\bf S}\cdot(\hat{\bf r}\times{\bf v}))\hat{\bf r}\right]-  \cr
\frac{1}{2mc}{\boldsymbol{\nabla}}([{\bf B}_J+{\bf B}_S]\cdot {\bf S}). \qquad \qquad \qquad \qquad
\end{eqnarray}
The first and second lines in (\ref{sh11}) come from the first term of the effective Hamiltonian (\ref{sh9}), while the
last line comes from the second term of (\ref{sh9}). The new term due to gravimagnetic moment is
$-\frac{1}{2mc}{\boldsymbol{\nabla}}({\bf B}_S \cdot {\bf S})$. As it should be expected, other terms coincide with
those of known from analysis of MPTD equations \cite{Einstein1915, Einstein1916, Thirring1918.2, Lense1918, Wald1972,
Mashhoon1982, deSitter1916, Adler2015, Thorne1998, Schiff1960.1, Connell1975}.  The first term in
(\ref{sh11}) represents the standard limit of Newtonian gravity and implies an elliptical orbit. The next three terms
represent an acceleration in the plane of orbit and are responsible for the precession of perihelia \cite{Einstein1915,
Einstein1916, Weinberg}. The  term $\frac{1}{c}{\bf B}_J\times{\bf v}$ represents the acceleration due to
Lense-Thirring rotation of central body \cite{Lense1918, Thirring1918.2, Mashhoon1982}, while the remaining terms
describe the influence of the gyroscopes spins on its trajectory. The gravitational dipole-dipole force
$\frac{1}{2mc}{\boldsymbol{\nabla}}({\bf B}_J\cdot {\bf S})$ has been computed by Wald \cite{Wald1972}. The new
contribution due to nonminimal interaction, $\frac{1}{2mc}{\boldsymbol{\nabla}}({\bf B}_S\cdot {\bf S})$, is similar
to the Wald term and is of the same (or less) magnitude.

In a co-moving frame, the effective Hamiltonian (\ref{sh9}) implies precession of spin $\frac{d{\bf
S}}{dt}=[{\boldsymbol\Omega}\times{\bf S}]$ with angular velocity vector
\begin{eqnarray}\label{sh12}
{\boldsymbol\Omega}=\frac{3GM}{2c^2r^2}[\hat{\bf r}\times{\bf v}]+\frac{1}{2c}{\bf B}_J+\frac{1}{c}{\bf B}_S.
\end{eqnarray}
The geodetic precession (first term in (\ref{sh12})) comes from the first term of effective Hamiltonian (\ref{sh9}), while
the frame-dragging precession (second term in (\ref{sh12})) is produced by the term $\frac{1}{2c}({\bf B}_J\cdot{\bf
S})$. So they are the same for both gravimagnetic and MPTD particle. They have been computed by Schiff
\cite{Schiff1960.1}, and measured during  the Stanford Gravity Probe B experiment \cite{GravityPB2015}. The last term in
(\ref{sh12})  appears only for  the gravimagnetic particle and depends on gyroscopes spin ${\bf S}$. Hence, two gyroscopes
with different magnitudes and  directions of spin will precess around different rotation axes. Then the angle between
their own rotation axes will change with time in Schwarzschild or Kerr space-time. Since the variation of the angle can be
measured with high precision, this effect could be used to find out whether a rotating body has unit or null
gravimagnetic moment.

Comparing the last two terms in (\ref{sh12}), we conclude that the precession of spin ${\bf S}$ due to gravimagnetic moment
is equivalent to that of caused by rotation of the central body with momentum ${\bf J}_{fict}=\frac{M}{m}{\bf S}$.

Effective Hamiltonian for the case of  the non rotating central body (Schwarzschild metric) is obtained from (\ref{sh9}) by
setting ${\bf A}_J={\bf B}_J=0$. We conclude that, due to the term $\frac{1}{2c}{\bf B}_S\cdot{\bf S}$,  spin of
gravimagnetic particle will experience frame-dragging precession with angular velocity $\frac{1}{c}{\bf B}_S$ even in the field of
non rotating central body, see (\ref{sh12}).

To estimate the relative magnitude of spin torques due to ${\bf B}_J$ and ${\bf B}_S$, we represent them in terms of
angular velocities. Assuming that the two bodies are spinning spheres of uniform density, we write
$\mathbf{J}=I_1{\boldsymbol{\omega}}_1$ and $\mathbf{S}= I_2{\boldsymbol{\omega}}_2$, where ${\boldsymbol{\omega}_i}$
is angular velocity and $I_i=(2/5)m_ir^2_i$ is  the moment of inertia. Then the last two terms in (\ref{sh12}) read
\begin{equation}\label{O-total2}
\mathbf{\Omega}_{fd}= \frac{2Gm_1r^2_1}{5c^2r^3}\left[ 3 \left([{\boldsymbol{\omega}}_1 + \rho^2
{\boldsymbol{\omega}}_2]\cdot\mathbf{\hat r}\right)\mathbf{\hat r} - ({\boldsymbol{\omega}}_1
+\rho^2{\boldsymbol{\omega}}_2)\right] \, ,
\end{equation}
where $\rho\equiv r_2/r_1$. Note that $\mathbf{\Omega}_{fd}$  does not  depend on mass of the test particle. The ratio
$\rho^2$ is extremely small for the case of Gravity Probe B experiment, so the MPTD and gravimagnetic bodies are
indistinguishable in this experiment.  For a system like Sun-Mercury $\rho^2\sim 10^{-5}$. For a system like
Sun-Jupiter  $\rho^2\sim 10^{-2}$. The two torques could have a comparable magnitudes in a binary system with stars of
the same size \cite{Manchester:2010dh, Homan:2012us,  Weisberg:2002qg} (then $\rho=1$), but one of them much heavier
than the other (neutron star or white dwarf). Then our approximation of a central field is reasonable, and according to
Eq. (\ref{O-total2}), the frame-dragging effect due to gravimagnetic moment  becomes  comparable  with the Schiff
frame-dragging effect.

To compare the two effects in a binary system with arbitrary masses, we need to go beyond the central-field
approximation. Probably, this case can be reduced to the central-field approximation following the procedure
\cite{Robertson1938, Connell1975, Will1993}.

{\bf Acknowledgments} We thank Wenbin Lin for helpful discussions.  The work of AAD has been supported by the Brazilian foundation CNPq (Conselho Nacional de
Desenvolvimento Cient\'ifico e Tecnol\'ogico - Brasil). WGR thanks CAPES for the financial support (Programm
PNPD/2011).


\end{document}